\documentclass{nrc2}

\journalcode{cjp}

\usepackage{graphicx}
\usepackage{amssymb}
\usepackage{url}
\usepackage{flushend}

\def\ao{{a_0}}
\def\ac{{a_c}}
\def\gn{{g_N}}
\def\vc{{v_c}}
\def\Mo{{M_0}}
\def\Md{{M_{\rm dyn}}}
\def\Mv{{M_{\rm vir}}}
\def\rv{{r_{\rm vir}}}
\def\mss{{m\,s$^{-2}$}}

\newcommand\vol[1]{{\bf #1}}
\newcommand\name[1]{{\small\sc #1}}

\title{The ``Graviton Picture'': a Bohr Model for Gravitation on Galactic Scales?}
\author{Sascha Trippe}
\address{Seoul National University, Department of Physics and Astronomy, Seoul 151-742, South Korea}
\correspond{trippe@astro.snu.ac.kr}
\shortauthor{S. Trippe}

\received{March 28, 2014}
\accepted{May 5, 2014}

\begin{document}

\begin{abstract}
Modified Newtonian Dynamics (MOND) provides a successful description of stellar and galactic dynamics on almost all astronomical scales. A key feature of MOND is the transition function from Newtonian to modified dynamics which corresponds to the empirical mass discrepancy--acceleration (MDA) relation. However, the functional form of the MDA relation does not follow from theory in a straightforward manner; in general, empirical MDA relations are inserted ad hoc into analyses of stellar dynamics. I revisit the possibility of gravity being mediated by massive virtual particles, \emph{gravitons}. Under certain reasonable assumptions, the resulting ``graviton picture'' implies a MDA relation that is equivalent to the -- empirical -- ``simple $\mu$'' function of MOND which is in very good agreement with observations. I conclude that the ``graviton picture'' offers a simple description of gravitation on galactic scales, potentially playing a role for gravitation analogous to the role played by Bohr's model for atomic physics.
\end{abstract}

\begin{resume}
Modified Newtonian Dynamics (MOND) provides a successful description of stellar and galactic dynamics on almost all astronomical scales. A key feature of MOND is the transition function from Newtonian to modified dynamics which corresponds to the empirical mass discrepancy--acceleration (MDA) relation. However, the functional form of the MDA relation does not follow from theory in a straightforward manner; in general, empirical MDA relations are inserted ad hoc into analyses of stellar dynamics. I revisit the possibility of gravity being mediated by massive virtual particles, \emph{gravitons}. Under certain reasonable assumptions, the resulting ``graviton picture'' implies a MDA relation that is equivalent to the -- empirical -- ``simple $\mu$'' function of MOND which is in very good agreement with observations. I conclude that the ``graviton picture'' offers a simple description of gravitation on galactic scales, potentially playing a role for gravitation analogous to the role played by Bohr's model for atomic physics.
\end{resume}

\maketitle

\section{Preamble \label{sec:intro}}

Dark matter is dead, to begin with -- at least in the sense of ``non-baryonic dark matter comprising 85\% of the dynamical mass of the universe''. A vast amount of observations collected during, especially, the last decade has enforced the conclusion that a solution of the ``missing mass problem'' of astronomy requires a modified law of gravity \cite{sanders2002,rhee2004a,rhee2004b,mcgaugh2004,mcgaugh2005a,mcgaugh2005b,famaey2012,kroupa2012,trippe2014}. Early proposals aimed at modifications of Newtonian gravity beyond characteristic length scales of few kiloparsecs \cite{finzi1963,sanders1984} were unsuccessful eventually. A breakthrough was achieved by Milgrom's Modified Newtonian Dynamics (MOND) \cite{milgrom1983a,milgrom1983b,milgrom1983c} which postulates a modification of Newtonian gravity -- with Newtonian gravitational acceleration $\gn$ -- as function of acceleration (gravitational field strength) $g$. This modification takes the form
\begin{equation}
\label{eq:mond}
\gn = \mu(x)\,g
\end{equation}
with $x=g/\ao$, $\ao$ being Milgrom's constant (of the dimension of an acceleration), and $\mu(x)$ denoting a transition function with the asymptotic behavior $\mu(x)=1$ for $x\gg1$ and $\mu(x)=x$ for $x\ll1$. The first limiting case corresponds to the usual Newtonian dynamics. The second limiting case -- i.e., $g\ll\ao$ -- leads to
\begin{equation}
\label{eq:vc}
\vc^4 = G\,\Mo\,\ao = const
\end{equation}
for stellar systems dominated by rotation (like disk galaxies) and
\begin{equation}
\label{eq:sigma}
\sigma^4 = \frac{4}{9}\,G\,\Mo\,\ao = const
\end{equation}
for stellar systems dominated by random motions (like elliptical galaxies or galaxy clusters) \cite{milgrom1984,milgrom1994}; here, $\vc$ denotes the circular speed of a particle around a luminous mass $\Mo$, $\sigma$ is the three-dimensional velocity dispersion, and $G$ is Newton's constant.\footnote{For simplicity, I only quote the absolute values of positions, velocities, and accelerations throughout this paper.} The scaling laws Eqs.~\ref{eq:vc}, \ref{eq:sigma} are in excellent agreement with observations, specifically the baryonic Tully--Fisher \cite{rhee2004a,rhee2004b,mcgaugh2005a,tully1977,mcgaugh2011} and Faber--Jackson \cite{faber1976,sanders2010} relations as well as the surface density--acceleration relation of disk galaxies \cite{mcgaugh2005b,famaey2012}; these observations constrain Milgrom's constant to a value of $\ao=(1.2\pm0.2)\times10^{-10}$\,\mss.

Whereas the \emph{asymptotic} scaling relations (Eqs.~\ref{eq:vc}, \ref{eq:sigma}) are well explored, the \emph{transitional} regime between Newtonian and modified dynamics, quantified by the function $\mu(x)$, is not -- at least not theoretically. The functional form of $\mu(x)$ does not follow from theory, like Bekenstein's Tensor--Vector--Scalar (TeVeS) theory \cite{bekenstein2004,bekenstein2006}, in a straightforward manner; instead, various functional relations have been assumed ad hoc for the analysis of stellar dynamics \cite{milgrom1983b,famaey2005}. A notable exception is provided by the analysis of \cite{milgrom1999}: assuming that MOND is the result of a modification of inertia due to vacuum effects -- specifically the Unruh effect -- it is possible to derive the expression $\mu(x)=x/(1+x^2)^{1/2}$ which is consistent with observations \cite{kroupa2012}. Empirically, $\mu(x)$ is constrained the strongest by the mass discrepancy--acceleration (MDA) relation of disk galaxies, meaning a strong anti-correlation between gravitational field strength $g$ and the ratio $\Md/\Mo$, with $\Md$ and $\Mo$ denoting dynamical\footnote{Referring to the mass required to explain the dynamics of a stellar system (a star cluster, a galaxy, a galaxy cluster, and so on) when assuming the validity of Newtonian gravity on astronomical scales.} and luminous\footnote{Referring to the mass stored in stars and interstellar or intergalactic gas that is actually observed, usually via optical, radio, and/or X-ray imaging.} mass, respectively \cite{sanders2002,mcgaugh2004,famaey2012}. Any theory of modified gravity which provides a prediction of the functional form of $\mu(x)$ -- and, eventually, any complete theory should do so -- can be tested by comparison to the empirical MDA relation in a straightforward manner.

In this article I revisit the ``graviton picture'', a model of gravity which employs the ad-hoc assumption that gravitation is mediated by virtual particles -- gravitons -- with non-zero mass \cite{trippe2013a}. This model provides (i) the usual MOND scaling laws (Eqs.~\ref{eq:vc}, \ref{eq:sigma}) in the limit $g\ll\ao$ \cite{trippe2013b} and (ii) a theoretical MDA relation in good agreement with the observed one \cite{trippe2013c}.

\section{The ``Graviton Picture'' \label{sec:gravitons}}

\subsection{Motivation \label{sec:motivation}}

The construction of any model of modified gravity begins with a census of the essential boundary conditions that need to be incorporated; in the following, I regard a simple dynamical system composed of a luminous point-like mass $\Mo$ orbited by a quasi-massless test particle with circular speed $\vc$ on a circular orbit with radius $r$ for reference. We can start off from two fundamental observations: (i) asymptotically flat rotation curves \cite{rubin1980}, implying\footnote{Using the usual Newtonian relation $\vc(R)^2=GM(R)/R$ with radial coordinate $R$ and mass profile $M(R)\propto R$ leads to $\vc=const$; this is the case if, and only if, the mass density obeys $\rho(R)\propto R^{-2}$.} distributions of the ``missing'' mass like $\rho(R)\propto R^{-2}$, with $\rho(R)$ being the mass density as function of radial coordinate $R$; and (ii) the observation that the luminous mass is a proxy for the dynamical mass in all dynamical systems and at all scales \cite{sancisi2004,swaters2012}. The latter condition demands that we find a model wherein luminous and dynamical mass -- and thus the density of the ``missing mass'' -- are related one-to-one. The former condition implies that the density follows an inverse-square-of-distance law. In combination, the two conditions lead to an interesting idea: \emph{the extra mass behaves as if it were a radiation composed of massive particles emitted isotropically from a source mass $\Mo$.}

With this idea being formulated, we immediately see that those particles must be of a peculiar nature. Firstly, the particle radiation must originate from (or couple to) the source mass $\Mo$ without removing mass (or energy) from it -- on the contrary, mass needs to be \emph{added} to $\Mo$. Secondly, these particles must be electromagnetically dark (or ``invisible'') in order to escape observation. Fortunately, a candidate is available: the \emph{graviton}, the virtual mediator of gravitational interaction in quantum field theories \cite{rosenfeld1930,rovelli2001,griffiths2008}. I note that, usually, gravitons are assumed to be massless, but this is not necessarily the case: consistent theoretical descriptions of gravity based on gravitons with non-zero mass can be, and have been, developed \cite{fierz1939,goldhaber2010,hinterbichler2012}.

\subsection{Fundamental Assumptions \label{sec:assumptions}}

Following up on the idea motivated in \S\ref{sec:motivation}, I make the following fundamental assumptions (A).

\begin{itemize}

\item[A1.] Gravitation is mediated by discrete exchange particles: gravitons.

\item[A2.] Gravitons are virtual particles arising from quantum fluctuations.

\item[A3.] Gravitons have a non-zero mass.

\item[A4.] Gravitational interactions are possible only between (i) two real masses, and (ii) a real mass and a graviton emitted by a real mass. Especially, graviton--graviton interactions are forbidden.

\end{itemize}

\subsection{Consequences \label{sec:consequences}}

\begin{figure*}[t!]
\centering
\includegraphics[angle=-90,width=175mm]{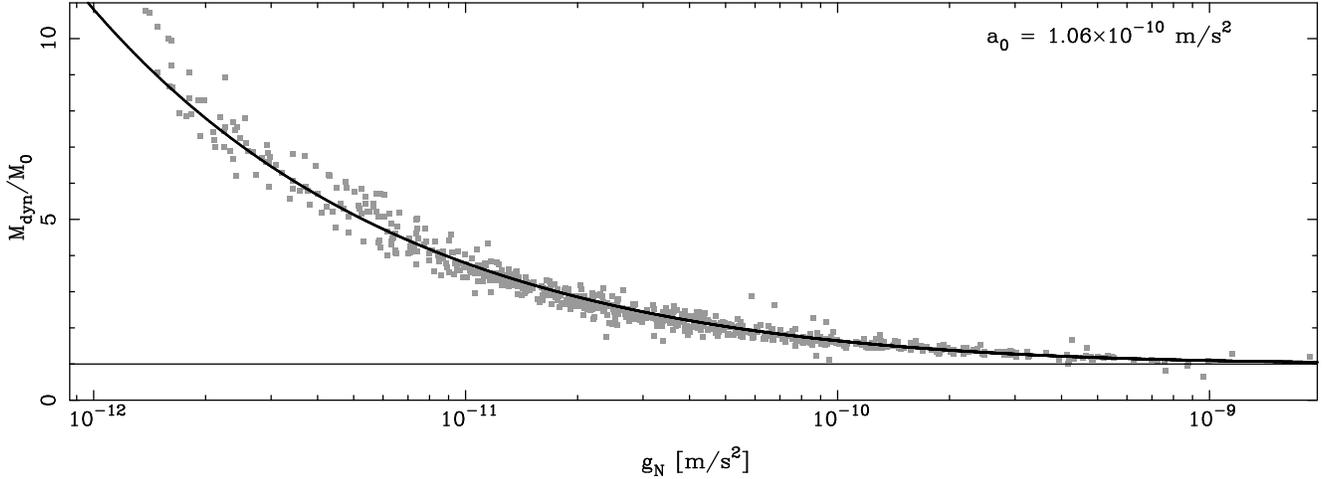}
\caption{Mass discrepancy $\Md/\Mo$ as function of Newtonian gravitational field strength $\gn$. Grey points denote observational data, in total 735 measurements from 60 galaxies \cite{famaey2012}. The continuous black curve is the law given by Eq.~\ref{eq:mda} for $\ao=1.06\times10^{-10}$\,\mss. \label{fig:mda}}
\end{figure*}

\subsubsection{Dynamical Mass \label{sec:dynmass}}

The assumptions A1--4 imply the following geometry: the mass $\Mo$ is the source of virtual, massive gravitons which are radiated away isotropically. These gravitons form a spherical halo around $\Mo$ with a density profile
\begin{equation}
\label{eq:rho}
\rho(R) = \beta\,\Mo\,R^{-2}
\end{equation}
with $\beta$ being a factor of the dimension of an inverse length. The proportionality to $\Mo$ follows from consistency with classical field theories which demand that a field strength is proportional to the source charge, with mass being the charge of gravity in our case. The proportionality to $R^{-2}$ mirrors the inverse-square-of-distance law of radiation. When integrating $\rho(R)$ over $R$ from 0 to $r$ while demanding that $\beta$ is not a function of the radial coordinate, the total, dynamical mass $\Md(r)$ is given by the sum of the initial source mass and the graviton mass enclosed within $r$,
\begin{equation}
\label{eq:mdyn1}
\Md = \Mo \left(1 + 4\pi\,\beta\,r\right) ~~ .
\end{equation}
At this point, we ought to make a choice for the parameter $\beta$ which is nothing else than the inverse of a characteristic radius. For simple, circular, dynamical systems as introduced in \S\ref{sec:motivation}, a characteristic radius is given by the kinetic energy per unit mass of the test particle, $\vc^2/2$, in units of a constant of the dimension of an acceleration, $\alpha$, thus suggesting the choice
\begin{equation}
\label{eq:beta}
\beta = \frac{2\alpha}{\vc^2} ~~ .
\end{equation}
Using the definition of the centripetal acceleration, $\ac=\vc^2/r$, we eventually find
\begin{equation}
\label{eq:mdyn2}
\Md = \Mo \left(1 + 8\pi\frac{\alpha}{\ac}\right)
\end{equation}
where the deviation from the identity of luminous and dynamical mass (i.e., the second summand) scales with the inverse of the centripetal acceleration.

\subsubsection{Graviton Mass \label{sec:gravitonmass}}

The assumption of gravitons having non-zero mass (A3) implies that this mass must be very small. Heisenberg's uncertainty relation \cite{heisenberg1927,kennard1927} for energy and time demands that the graviton mass must be small enough to permit a graviton life time on the order of the Hubble time $T_H\approx1.4\times10^{10}$ years -- meaning a mass of less than roughly $10^{-69}$\,kg or $10^{-33}$\,eV\,$c^{-2}$. Due to the vast differences in particle life times, these values are about 40 orders of magnitude lower than the masses of the exchange particles mediating the nuclear forces \cite{griffiths2008}.

\section{Discussion \label{sec:discussion}}

\subsection{Consistency with MOND \label{sec:mondlimits}}

Evidently, any modified law of gravitation -- Eq.~\ref{eq:mdyn2} in our case -- has to comprise the relevant limits: Newtonian dynamics for $\ac\gg\alpha$, and modified Newtonian dynamics for $\ac\ll\alpha$. The former case is self-evident: for $\ac\gg\alpha$, the second term in Eq.~\ref{eq:mdyn2} vanishes, leading to $\Md=\Mo$ and thus removing any mass discrepancy.

In the case $\ac\ll\alpha$, the second term in Eq.~\ref{eq:mdyn2} becomes dominant, meaning $\Md\approx8\pi\alpha/\ac$. Using the relations $\vc^2=G\Md/r$ and $\ac=\vc^2/r$, we can bring this expression into the form
\begin{equation}
\label{eq:vcgrav1}
\vc^4 = 8\,\pi\,G\,\Mo\,\alpha ~~ .
\end{equation}
Comparison of Eqs.~\ref{eq:vc} and \ref{eq:vcgrav1} shows that the expressions are equivalent, with Milgrom's constant $\ao$ and the constant $\alpha$ being related like $\ao=8\pi\alpha$.

\subsection{Agreement with the MDA Relation \label{sec:mda}}

The transitional regime between Newtonian and modified Newtonian dynamics is quantified by the evolution of the ratio of dynamical and luminous mass, $\Md/\Mo$, as function of gravitational acceleration ($g$ or $\ac$) -- the \emph{mass discrepancy--acceleration relation}. The best dynamical probe is provided by disk galaxies: flattened, rotating stellar systems with the rotation dominating over random motions. Dynamical masses can be derived from the stellar rotation speeds as function of galactocentric radius, luminous masses are derived from the amount of light emitted by stars and interstellar gas. Disk galaxies show \cite{sanders2002,mcgaugh2004,famaey2012} a strong and characteristic anti-correlation between the ratio $\Md/\Mo$ and gravitational acceleration. This anti-correlation is universal -- all data from all galaxies studied fall on the same curve (within measurement uncertainties). I illustrate this behavior in Fig.~\ref{fig:mda} which displays $\Md/\Mo$ as function of the Newtonian gravitational acceleration $\gn = G\Mo/r^2 = g(\Mo/\Md)$, using data from \cite{famaey2012} after correcting for a systematic offset \cite{trippe2013c}.

Using the relation $\ao=8\pi\alpha$, we can re-write Eq.~\ref{eq:mdyn2} as
\begin{equation}
\label{eq:mda}
\frac{\Md}{\Mo} = 1 + \frac{\ao}{\ac} ~ .
\end{equation}
Noting that $\ac=g$ for circular orbits and that $\Md/\Mo=g/\gn$, we realize that the mass discrepancy as function of acceleration is simply the inverse of $\mu(x)$ in Eq.~\ref{eq:mond}. Indeed, the relation given by Eq.~\ref{eq:mda} is equivalent to \cite{trippe2013b} the empirical ``simple $\mu$ function'' of MOND used for modeling galactic rotation curves \cite{famaey2005} as well as the MDA relation \cite{kroupa2012}. I demonstrate the good agreement between Eq.~\ref{eq:mda} and the MDA data in Fig.~\ref{fig:mda}, with the best-fit value for Milgrom's constant being $\ao=(1.06\pm0.05)\times10^{-10}$\,\mss\ \cite{trippe2013c}.

\subsection{Virial Radii \label{sec:virial}}

As noted by \cite{wu2013}, an effective dynamical mass $\Md$ like the one given by Eq.~\ref{eq:mda} can be identified with the \emph{virial mass} of a stellar system -- comprising luminous mass plus apparent phantom dark matter halo -- like
\begin{equation}
\label{eq:virialmass}
\Md(\rv) \equiv \Mv = p\rv^3
\end{equation}
with $p=(4/3)\pi\times200\rho_{\rm crit}$, $\rho_{\rm crit}=3H_0^2/(8\pi G)$ being the critical density of the universe, and $H_0\approx70$\,km\,s$^{-1}$\,Mpc$^{-1}$ being Hubble's constant (e.g., \cite{lee2012}). By construction, the \emph{virial radius} $\rv$ is the radius enclosing a dynamical mass with an average density which is 200 times the critical density of the universe. Comparison of Eqs. \ref{eq:mda} and \ref{eq:virialmass} for the case $\ac\ll\ao$ leads to a virial radius of
\begin{equation}
\label{eq:virialradius}
\rv = \left(\frac{\Mo\,\ao}{p\,\vc^2}\right)^{1/2} = \left(\frac{\Mo\,\ao}{p^2\,G}\right)^{1/4}
\end{equation}
for $\ac(r)\equiv\ac(\rv)=\vc^2/\rv$, with the second equality following from Eq.~\ref{eq:vc}. This may be compared to the Newtonian expression $\rv=\vc/(pG)^{1/2}$.

\subsection{Assumptions on Gravitons \label{sec:gravitonassumptions}}

The assumption (A3) that gravity is mediated by massive particles (also referred to as \emph{massive gravity}) with limited life times is in tension with the standard assumption that the range of gravity is infinite. Gravitational fields mediated by massive gravitons decay exponentially, resulting in a Yukawa profile \cite{hinterbichler2012}. Graviton masses consistent with graviton life times (the decay time scale) on the order of the Hubble time $T_H$ provide for a decay of gravity on cosmological scales and an apparent acceleration of the expansion of the universe -- as actually observed. Indeed, recent studies conclude that cosmological observations are in agreement with cosmological models based on massive gravity \cite{volkov2012,cardone2012}.

For reasons of self-consistency, I introduced the assumption (A4) that gravitons do not interact with each other. A priori, mediator particles which themselves carry the charge they couple to -- mass in our case -- should also interact with each other, leading to substantial deviations from the classical force laws. A good example are the color charges of gluons in quantum chromodynamics which lead to the confinement of quarks \cite{griffiths2008,wilson1974}. Notably, the assumption that massive gravitons do not interact with each other is a common feature of massive gravity theories \cite{goldhaber2010,hinterbichler2012}.

\subsection{Limitations \label{sec:limitations}}

Despite its success in describing galactic dynamics, the ``graviton picture'' is subjected to serious limitations. Firstly, it is, a priori, non-relativistic; the scaling relations I derive from the ``graviton picture'' (Eqs. \ref{eq:mdyn2}, \ref{eq:mda}) follow from classical field theory and classical mechanics. Furthermore, the assumptions A1--4 are, although motivated by ongoing research efforts, ad hoc; their consistency with quantum field theories is unclear.

Secondly, the scaling law given by Eq.~\ref{eq:mda} is probably\footnote{As noted explicitly by \cite{anderson1995}, the accuracy of the limits on ``missing mass'' in the solar system is affected by (i) non-obvious correlations between errors, and (ii) the presence of the Kuiper belt in case of Neptune.} inconsistent with the kinematics of the solar system. The relative ``extra mass'' added to the mass of the sun, i.e., $\Md/\Mo-1$, is, in units of $10^{-7}$, 5, 66, and 162 for Jupiter, Uranus, and Neptune, respectively (using orbit data from \cite{tholen2000} and assuming $\ao=1.06\times10^{-10}$\,\mss). The observational limits ($3\sigma$ upper limits) are 2, 18, and 35, respectively \cite{anderson1995} -- meaning that the predicted values exceed the observational limits by factors up to about five.

The concerns raised above indicate a failure of the ``graviton picture'' in the regime of strong -- relative to $\ao$ -- gravitational fields. Remarkably, the discrepancy between the predictions of Eq.~\ref{eq:mda} and solar system kinematics is, although highly significant, still very small in absolute terms: in a regime where $g/\ao\sim10^{5}$, the discrepancy between Eq.~\ref{eq:mda} and the observational limits is $(\Md/\Mo)-1\lesssim10^{-5}$. Accordingly, the ``graviton picture'' provides a good approximation formula for describing the dynamics in gravitational fields with $\log_{10}(g/\ao)$ ranging approximately from $-1$ (in the outskirts of galaxies -- cf. Fig.~\ref{fig:mda}) to $+5$ (in the outer solar system) at the very least.

\section{Epilogue \label{sec:summary}}

I revisited the ``graviton picture'' of gravitation, a model based on the ad-hoc assumption that gravitation is mediated by massive gravitons that obey certain reasonable rules of interaction. This model provides a scaling law for the ratio of dynamical and luminous mass which is in good agreement with the empirical mass discrepancy--acceleration of disk galaxies. More generally speaking, this suggests that if gravity is mediated by a massive boson in a way that preserves the limits of classical field theory, the resulting potential is MONDian. Eventually, I arrive at two principal conclusions.

Firstly, the MDA relation is, arguably, the most important test for any theory of gravity in the regime of weak gravitational fields. Indeed, the asymptotic relations given by Eqs.~\ref{eq:vc} and \ref{eq:sigma} derive from \emph{any} law of gravity scaling with acceleration (field strength) -- meaning they are actually unable to discriminate between theories (e.g., \cite{famaey2012}). Any complete theory must provide for a theoretical MDA relation which in turn must agree with the MDA data.

Secondly, the ``graviton picture'' is, despite its success on galactic scales, not yet a consistent theory of gravity. However, it bears strong similarity to one of the most famous heuristic models of physics: Bohr's ``planetary system'' model of the structure of atoms \cite{bohr1913a,bohr1913b}. Even though it is obvious today that atoms are not even remotely similar to microscopic planetary systems, Bohr's model provided the first successful quantitative prediction of the spectrum of hydrogen and hydrogen-like atoms -- an approximation satisfactory for many purposes until the present day. A similar role might, eventually, be played by the ``graviton picture'' for gravity: despite its incompleteness, the ``graviton picture'' provides a successful prediction of the empirical MDA relation and thus for the dynamics of gravitationally bound systems spanning interstellar and galactic scales.

\section*{Acknowledgments}

This work made use of the galactic dynamics data base provided by \name{Stacy S. McGaugh} at Case Western Reserve University, Cleveland (Ohio, U.S.A.),\footnote{\url{http://astroweb.case.edu/ssm/data/}} and of the software package \name{dpuser} developed and maintained by \name{Thomas Ott} at MPE Garching (Germany).\footnote{\url{http://www.mpe.mpg.de/~ott/dpuser/dpuser.html}} I acknowledge financial support from the Korean National Research Foundation (NRF) via Basic Research Grant no. 2012-R1A1A2041387. Last but not least, I am grateful to an anonymous referee for valuable comments.


\begin{thebibliography}{99}

\bibitem{sanders2002} R.H. Sanders and S.S. McGaugh, Annu. Rev. Astron. Astrophys. \vol{40}, 263 (2002)
\bibitem{rhee2004a} M.-H. Rhee, J. Kor. Astron. Soc. \vol{37}, 15 (2004)
\bibitem{rhee2004b} M.-H. Rhee, J. Kor. Astron. Soc. \vol{37}, 91 (2004)
\bibitem{mcgaugh2004} S.S. McGaugh, Astrophys. J. \vol{609}, 652 (2004)
\bibitem{mcgaugh2005a} S.S. McGaugh, Astrophys. J. \vol{632}, 859 (2005)
\bibitem{mcgaugh2005b} S.S. McGaugh, Phys. Rev. Lett. \vol{95}, 171302 (2005)
\bibitem{famaey2012} B. Famaey and S.S. McGaugh, Living Rev. Relativ. \vol{15}, 10 (2012)
\bibitem{kroupa2012} P. Kroupa, Publ. Astron. Soc. Aust. \vol{29}, 395 (2012)
\bibitem{trippe2014} S. Trippe, Z. Naturforsch. A \vol{69}, 173 (2014)
\bibitem{finzi1963} A. Finzi, Mon. Not. R. Astron. Soc. \vol{127}, 21 (1963)
\bibitem{sanders1984} R.H. Sanders, Astron. Astrophys. \vol{136}, L21 (1984)
\bibitem{milgrom1983a} M. Milgrom, Astrophys. J. \vol{270}, 365 (1983)
\bibitem{milgrom1983b} M. Milgrom, Astrophys. J. \vol{270}, 371 (1983)
\bibitem{milgrom1983c} M. Milgrom, Astrophys. J. \vol{270}, 384 (1983)
\bibitem{milgrom1984} M. Milgrom, Astrophys. J. \vol{287}, 571 (1984)
\bibitem{milgrom1994} M. Milgrom, Astrophys. J. \vol{429}, 540 (1994)
\bibitem{tully1977} R.B. Tully and J.R. Fisher, Astron. Astrophys. \vol{54}, 661 (1977)
\bibitem{mcgaugh2011} S.S. McGaugh, Phys. Rev. Lett. \vol{106}, 121303 (2011)
\bibitem{faber1976} S.M. Faber and R.E. Jackson, Astrophys. J. \vol{204}, 668 (1976)
\bibitem{sanders2010} R.H. Sanders, Mon. Not. R. Astron. Soc. \vol{407}, 1128 (2010)
\bibitem{bekenstein2004} J.D. Bekenstein, Phys. Rev. D \vol{70}, 083509 (2004)
\bibitem{bekenstein2006} J.D. Bekenstein, Contemp. Phys. \vol{47}, 387 (2006)
\bibitem{famaey2005} B. Famaey and J. Binney, Mon. Not. R. Astron. Soc. \vol{363}, 603 (2005)
\bibitem{milgrom1999} M. Milgrom, Phys. Lett. A \vol{253}, 273 (1999)
\bibitem{trippe2013a} S. Trippe, J. Kor. Astron. Soc. \vol{46}, 41 (2013)
\bibitem{trippe2013b} S. Trippe, J. Kor. Astron. Soc. \vol{46}, 93 (2013)
\bibitem{trippe2013c} S. Trippe, J. Kor. Astron. Soc. \vol{46}, 133 (2013)
\bibitem{rubin1980} U.C. Rubin, W.K. Ford, Jr. and N. Thonnard, Astrophys. J. \vol{238}, 471 (1980)
\bibitem{sancisi2004} R. Sancisi, in: S.D. Ryder et al. (eds.), IAU Symp. \vol{220}, 233 (2004)
\bibitem{swaters2012} R.A. Swaters et al., Mon. Not. R. Astron. Soc. \vol{425}, 2299 (2012)
\bibitem{rosenfeld1930} L. Rosenfeld, Ann. Phys. (Berlin) \vol{397}, 113 (1930)
\bibitem{rovelli2001} C. Rovelli, \texttt{arXiv:gr-qc/0006061} (2001)
\bibitem{griffiths2008} D. Griffiths, Introduction to Elementary Particles, 2nd edn., Wiley-VCH, Weinheim (2008)
\bibitem{fierz1939} M. Fierz and W. Pauli, Proc. R. Soc. London A \vol{173}, 211 (1939)
\bibitem{goldhaber2010} A.S. Goldhaber and M.M. Nieto, Rev. Mod. Phys. \vol{82}, 939 (2010)
\bibitem{hinterbichler2012} K. Hinterbichler, Rev. Mod. Phys. \vol{84}, 671 (2012)
\bibitem{heisenberg1927} W. Heisenberg, Z. Phys. \vol{43}, 172 (1927)
\bibitem{kennard1927} E.H. Kennard, Z. Phys. \vol{44}, 326 (1927)
\bibitem{wu2013} Wu, X. \& Kroupa, P., Mon. Not. R. Astron. Soc. \vol{435}, 1536 (2013)
\bibitem{lee2012} Lee, M.G. \& Jang, I.S., Astrophys. J. Lett. \vol{760}, L14 (2012)
\bibitem{volkov2012} M.S. Volkov, J. High Energy Phys. \vol{1}, 35 (2012)
\bibitem{cardone2012} V.F. Cardone, N. Radicella and L. Parisi, Phys. Rev. D \vol{85}, 124005 (2012)
\bibitem{wilson1974} K.G. Wilson, Phys. Rev. D \vol{10}, 2445 (1974)
\bibitem{tholen2000} D.J. Tholen, in: Allen's Astrophysical Quantities, ed. A.N. Cox, 4th. edn., New York: Springer (2000) 
\bibitem{anderson1995} J.D. Anderson et al., Astrophys. J. \vol{448}, 885 (1995)
\bibitem{bohr1913a} N. Bohr, Phil. Mag \vol{26}, 1 (1913)
\bibitem{bohr1913b} N. Bohr, Phil. Mag \vol{26}, 476 (1913)

\end{thebibliography}
\end{document}